\documentclass[cits]{PoS}
\usepackage{graphicx,amssymb,epsfig,epsf}
\usepackage{amsmath}
\usepackage{axodraw}

\boldmath
\title{The MSSM with a softly broken U$\bf{(2)^3}$ flavor symmetry}

\ShortTitle{The MSSM with a softly broken $U(2)^3$ flavor symmetry}
\unboldmath
\author{\speaker{Andreas CRIVELLIN}\\
        Institut f\"ur Theoretische Teilchenphysik,
        Universit\"at Bern,\\
        Ch-3004 Bern, Switzerland\\
        E-mail: \email{crivellin@itp.unibe.ch}}

\author{Lars HOFER\\
        Institut f\"ur Theoretische Physik und Astrophysik,
        Universit\"at W\"urzburg,\\
        D-97074 W\"urzburg, Germany\\
        E-mail: \email{lhofer@physik.uni-wuerzburg.de}}

\author{Ulrich NIERSTE\\
        Institut f\"ur Theoretische Teilchenphysik,
        Karlsruhe Institute of Technology,\\
        D-76128 Karlsruhe, Germany\\
        E-mail: \email{nierste@particle.uni-karlsruhe.de}}

\abstract{In this article we review the phenomenological consequences of radiative flavor-violation (RFV) in the MSSM. In the model under consideration the $U(3)^3$ flavor symmetry of the gauge sector is broken in a first step to $U(2)^3$ by the top and bottom Yukawa couplings of the superpotential (and possibly also by the bilinear SUSY-breaking terms). In a second step the remaining $U(2)^3$ flavor symmetry is softly broken by the trilinear $A$-terms in order to obtain the measured quark masses and the CKM matrix of the Standard Model (SM) at low energies.

The phenomenological implications of this model depend on the actual choice of the SUSY breaking $A$-terms. If the CKM matrix is generated in the down sector (by $A^d$), $B_s\to\mu^+\mu^-$ receives non-decoupling contributions from Higgs penguins which become important already for moderate values of $\tan\beta$. Also the $B_s-\bar{B}_s$ mixing amplitude can be significantly modified compared to the SM prediction including a potential induction of a new CP-violating phase (which is not possible in the MSSM with MFV).}

\FullConference{European Physical Society Europhysics Conference on High Energy Physics\\
                 July 21-27, 2011\\
                 Grenoble, France}

\begin{document}

\section{Introduction}
The most popular solution to the SUSY flavor problem is the
hypothesis of minimal flavor violation (MFV) \cite{MFV}: The soft
SUSY-breaking terms are assumed to preserve a $U(3)_Q\times U(3)_u\times
U(3)_d$ flavor symmetry, broken only by the Yukawa matrices $Y^{u\,(0)}$
and $Y^{d\,(0)}$. The imposed $U(3)^3$ symmetry is however in conflict
with hints on new CP violating phases in $B_d-\bar{B}_d$ and $B_s-\bar{B}_s$ mixing,
indicated by the D0 measurement of the dimuon charge asymmetry
\cite{Abazov:2011yk}.\footnote{Due to the recent LHCb measurement of
  the CP asymmetry in $B_s\to J/\psi \phi$ the situation is inconclusive
  at the moment. While the dimuon asymmetry measured by D0 points
  towards physics beyond the SM with $3.9\,\sigma$ 
significance and the CP
  asymmetry in $B_s\to J/\psi \phi$ measured by CDF has the same sign,
  LHCb obtained the opposite sign for the phase in $B_s-\bar{B}_s$ mixing
  (compatible with zero) from $B_s\to J/\psi \phi$.} This effect cannot
be explained within MSSM-MFV scenarios without enhancing
$B_s\to\mu^+\mu^-$ far above the current experimental bounds\footnote{In MFV
scenarios with an extended Higgs sectors, however, a large CP phase in
$B-\bar{B}$ mixing is possible \cite{Altm}.}. Moreover,
while a solution of the hierarchy problem favors stop masses well
  below 1 TeV, direct searches at the LHC resulted in tight bounds on
the masses of the squarks of the first two generations. This conflict is
a further challenge for the $U(3)^3$ symmetry.\medskip

These facts suggest to abandon the $U(3)^3$ flavor symmetry and to
settle for a $U(2)^3$ for the first two generations in order to avoid
conflicts with the tight constraints from Kaon and D-physics. A
corresponding relaxed MFV scenario with the Yukawas $Y^{u\,(0)}$ and
$Y^{d\,(0)}$ in the superpotential being the spurions breaking the
$U(2)^3$ has been studied in Ref.~\cite{U2MFV}. We consider in this
article a different scenario: We assume that the Yukawas
$Y^{u\,(0)}$ and $Y^{d\,(0)}$ preserve the $U(2)^3$ flavor symmetry and
that the soft SUSY-breaking trilinear $A$-terms $A^u$ and $A^d$ are the
spurions breaking it. Such a model is quite appealing because it links
the breaking of flavor symmetries to the breaking of supersymmetry. In
the quark sector the $U(2)^3$ symmetry is then only softly broken, in
the sense that the effective low-energy values of the Yukawa couplings
$Y^u_{\textrm{eff}}$ and $Y^d_{\textrm{eff}}$, which are linked to the
measured quark masses and CKM elements, are induced by $A^u$ and $A^d$
through radiative corrections. In this way the smallness of the light
quark masses is explained via loop-suppression
\cite{Borzumati:1999sp}\footnote{For the corresponding analysis in the
  lepton sector see Ref.~\cite{Crivellin:2010ty}}.\medskip

In this article we review our model of radiative flavor violation (RFV) and demonstrate that it can provide the above-mentioned new CP phase in $B_s-\bar{B}_s$ mixing in contrast to MFV. For a detailed study of further phenomenological consequences of the RFV scenario we refer to Ref.~\cite{RFV}.\bigskip

\section{Radiative flavor violation (RFV)}
In our scenario of RFV we assume that the bare Yukawa couplings $Y^{u\,(0)}$ and $Y^{d\,(0)}$ in the superpotential exhibit a $U(2)_Q\times U(2)_u\times U(2)_d$ flavor symmetry:
\begin{equation}
  Y^{q(0)}  = \left( {\begin{array}{*{20}c}
   0 & 0 & 0  \\
   0 & 0 & 0  \\
   0 & 0 & y^q  \\
\end{array}} \right),\hspace{1cm} (q=u,d).
\end{equation}
While the bilinear soft SUSY-breaking mass terms are assumed to possess the same symmetry,
the trilinear $A$-terms $A^{u}$ and $A^{d}$ are the spurions breaking it. We perform $U(2)$-rotations on
the left- and right-handed up- and down-quark superfields to fix the basis in flavor space such that
\begin{equation}
   A^q = \left( {\begin{array}{*{20}c}
   A^q_{11}&   0 & A^q_{13}  \\
   0 &   A^q_{22} & A^q_{23}  \\
   A^q_{31} & A^q_{32} & A^q_{33}  \\
\end{array}} \right),\hspace{1cm} (q=u,d).
\end{equation}
Note that the resulting basis is not a weak eigenbasis because left-handed up- and down-fields have to be rotated independently in order to diagonalize the $2\times 2$ blocks of $A^u$ and $A^d$ simultaneously. Since in this basis no sources of flavor-violating (1,2) elements are present in the squark-mass matrices, the corresponding
CKM matrix $V^{(0)}_{2\times 2}$ equals the Cabibbo matrix $V_{2\times 2}$
known from experiment (up to negligible corrections arising from loops
involving a $1\to3\to 2$ transition):
\begin{equation}
  V^{(0)}  = \left( {\begin{array}{*{20}c}
   \cos\theta_C &   \sin\theta_C & 0  \\
     -\sin\theta_C &   \cos\theta_C & 0  \\
   0 & 0 & 1  \\
\end{array}} \right).
\end{equation}\medskip

The $U(2)^3$-breaking in the squark sector leaks into the quark
sector via loop effects.  The measured quark masses of the first two
generations and the measured CKM elements are manifestations of the
$U(2)^3$-breaking and as a consequence they must be directly related to
the $A^{u,d}_{ij}$. Neglecting multiple flavor transitions (except for
$1\to 2\to 3$ transitions) and small quark mass ratios one has
\begin{eqnarray}
   m_{q_i}&=&a_q\,\dfrac{A^q_{ii}}{\mu_A}\,v_q\,,\hspace{2cm} (q=u,d,\hspace{0.5cm}i=1,2)\,,\nonumber\\
   V_{cb}\,\approx\,-V_{ts}^*&=&b_d\,\dfrac{A^d_{23}}{\mu_A}\,\dfrac{v_d}{m_b}\,-\,b_u\,\dfrac{A^u_{23}}{\mu_{A}}\,\dfrac{v_u}{m_t}\,,\nonumber\\
   V_{ub}&=&b_d\,\left(\dfrac{A^d_{13}}{\mu_A}\,+\,V_{us}\dfrac{A^d_{23}}{\mu_A}\right)\,\dfrac{v_d}{m_b}\,-\,b_u\,\dfrac{A^u_{13}}{\mu_A}\,\dfrac{v_u}{m_t}\,,\nonumber\\
   -V_{td}^*&=&b_d\,\dfrac{A_{13}^d}{\mu_A}\,\dfrac{v_d}{m_b}\,-\,b_u\,\left(\dfrac{A^u_{13}}{\mu_A}\,+\,V_{cd}^*\,\dfrac{A_{23}^u}{\mu_A}\right)\,\dfrac{v_u}{m_t}\,, \label{eq:SoftFlavorBreaking}
\end{eqnarray}
where $\mu_A={\cal O}(A_{ii}^q)$ is a redundant mass scale
  introduced to render $a_q$, $b_q$ dimensionless. 
The coefficients $a_q$, $b_q$ are obtained by explicit evaluation of the
self-energy diagrams inducing the quark mass terms. Restricting
ourselves to SUSY-QCD contributions we find at first order in the mass
insertion approximation
\begin{equation}
   a_{u,d}\,=\,-\dfrac{2\alpha_s}{3 \pi} m_{\tilde{g}} \,\mu_A\, C_0\left(m_{\tilde{g}}^2, m_{\tilde{q}_{L}}^2,m_{\tilde{u}_{R},\tilde{d}_{R}}^2 \right),\hspace{1cm}
   b_{u,d}\,=\,-\dfrac{2\alpha_s}{3 \pi} m_{\tilde{g}} \,\mu_A\, C_0\left(m_{\tilde{g}}^2, m_{\tilde{q}_{L}}^2,m_{\tilde{t}_{R},\tilde{b}_R}^2 \right).
\end{equation}
Here $m_{\tilde{q}_{L}}$, $m_{\tilde{u}_{R}}$ and $m_{\tilde{d}_{R}}$ denote the common mass of the first two generations of left- and right-handed up- and down-type squarks, respectivly. If one assumes a $U(3)^3$ flavor symmetry for the bilinear squark mass terms (rather than only $U(2)^3$), one has $a_q=b_q$.\medskip

Eq.~(\ref{eq:SoftFlavorBreaking}) implies that the $A$-terms $A_{13}^q$
and $A_{23}^q$ exhibit a similar hierarchy with respect to each other as
the CKM-elements $V_{ub}$ and $V_{cb}$, and in particular that
$A^q_{11}/A^q_{22}=m_{q_1}/m_{q_2}$. The overall smallness, on the other
hand, of the masses of the first two quark generations and of the
off-diagonal CKM elements $V_{3i,i3}$ is explained by the loop
suppression in $a_q$, $b_q$. The SUSY-flavor problem is restricted
to the quantities $A^q_{31,32}$ which are not constrained from the
measured CKM elements since their contributions are suppressed by small
quark mass ratios.\medskip

\section{Higgs (double) penguins: RFV vs. MFV}
\begin{figure}[t]
  \begin{picture}(200,60) (-170,-40)
    \SetColor{Black}
    \ArrowLine(-140,10)(-105,-10)
    \ArrowLine(-105,-10)(-140,-30)
    \DashLine(-105,-10)(-45,-10){5}
    \Text(-135,10)[lb]{\Black{$b_R$}}
    \Text(-135,-35)[lb]{\Black{$s_L$}}
    \SetColor{MidnightBlue}
    \Vertex(-105,-10){4}
    \Text(-108,-3)[lb]{\MidnightBlue{$\Gamma^{LR}_{sb}$}}
    \SetColor{Black}
    \ArrowLine(-45,-10)(-10,10)
    \ArrowLine(-10,-30)(-45,-10)
    \Text(-95,-25)[lb]{\Black{$h^0,H^0,A^0$}}
    \Text(-20,12)[lb]{\Black{$\mu^-$}}
    \Text(-20,-45)[lb]{\Black{$\mu^+$}}

    \SetColor{Black}
    \ArrowLine(60,10)(95,-10)
    \ArrowLine(95,-10)(60,-30)
    \DashLine(95,-10)(155,-10){5}
    \Text(65,10)[lb]{\Black{$b_R$}}
    \Text(65,-35)[lb]{\Black{$s_L$}}
    \SetColor{MidnightBlue}
    \Vertex(95,-10){4}
    \Text(92,-3)[lb]{\MidnightBlue{$\Gamma^{LR}_{sb}$}}
    \SetColor{Black}
    \ArrowLine(155,-10)(190,10)
    \ArrowLine(190,-30)(155,-10)
    \Text(105,-25)[lb]{\Black{$h^0,H^0,A^0$}}
    \Text(180,12)[lb]{\Black{$s_R$}}
    \Text(180,-40)[lb]{\Black{$b_L$}}
    \SetColor{MidnightBlue}
    \Vertex(155,-10){4}
    \Text(145,-3)[lb]{\MidnightBlue{$\Gamma^{LR\,*}_{bs}$}}
    \SetColor{Black}
  \end{picture}
\caption{Higgs (double) penguin contributions to $B_s\to\mu^+\mu^-$ and to $B_s-\bar{B}_s$ mixing}
\label{fig:HiggsPeng}
\end{figure}
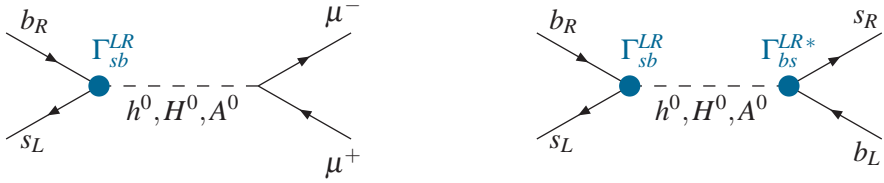

Loop diagrams involving the trilinear $A$-terms induce flavor-changing
neutral Higgs-quark couplings
$\Gamma^{H^0_k\;LR}_{q_iq_j}\,=\,\Gamma^{LR}_{q_iq_j}\,x_q^k$ ($q=u,d$)
where $H^0_k=(H^0,h^0,A^0)$ \cite{Crivellin:2010er}. The couplings
$\Gamma^{LR}_{sb},\;\Gamma^{LR}_{bs}$ contribute via Higgs (double)
penguin diagrams to the decay $B_s\to\mu^+\mu^-$ and to $B_s-\bar{B}_s$
mixing (see Fig.~\ref{fig:HiggsPeng}).
For the relevant Wilson coefficients one has
\begin{eqnarray}
  B_s\to\mu^+\mu^-\,:&&\hspace{0.5cm}C_S\,\propto\,\Gamma^{LR}_{sb},\hspace{0.5cm} C_S^{\prime}\,\propto\,\Gamma^{LR\,*}_{bs};\nonumber\\
  B_s-\bar{B}_s\textrm{ mixing}\,:&&\hspace{0.5cm} C^{LR}_2\,\propto\, \Gamma^{LR}_{sb}\,\Gamma^{LR\,*}_{bs}.
\end{eqnarray}
Note that contributions to $B_s-\bar{B}_s$ mixing which are proportional
to $(\Gamma^{LR}_{sb})^2$ or $(\Gamma^{LR\,*}_{bs})^2$ are strongly
suppressed and thus negligible. This is a consequence of a Peccei-Quinn
symmetry obeyed by the tree-level Yukawa-Lagrangian and the tree-level
Higgs potential of the MSSM \cite{Gorbahn:2009pp}.\medskip

As the effective couplings $\Gamma^{LR}_{sb}$ and $\Gamma^{LR}_{bs}$
break the $U(2)^3$ flavor symmetry, they must be directly related to the
corresponding spurions. In the MFV scenario we thus have\footnote{Full
  expressions for the Higgs (double) penguin contributions to
  $B_s\to\mu^+\mu^-$ and $B_s-\bar{B}_s$ mixing in the MSSM with MFV
  including non-decoupling effects and complex phases for SUSY
  parameters can be found in Ref.~\cite{HNS}.}
\begin{equation}
   \Gamma^{LR}_{sb}\,\propto\,\left(Y^uY^{u\dagger}Y^d\right)_{23}\,\approx\,
   y_b\,y_t^2\,V_{ts}^*V_{tb}\,,\hspace{1cm}
   \Gamma^{RL}_{sb}\,=\,\Gamma^{LR*}_{bs}\,\propto\,\left(Y^uY^{u\dagger}Y^d\right)_{32}^*\,\approx\,
   y_s\,y_t^2\,V_{ts}^*V_{tb}\,.
\end{equation}
Therefore $\Gamma^{LR}_{bs}$ is suppressed compared to
$\Gamma^{LR}_{sb}$ by the small quark mass ratio $m_s/m_b$. Experimental
bounds on $B_s\to\mu^+\mu^-$ constrain $\Gamma^{LR}_{sb}$ and, because
of the suppression of $\Gamma^{LR}_{bs}$ with respect to
$\Gamma^{LR}_{sb}$, they render Higgs double penguin effects in
$B_s-\bar{B}_s$ mixing negligible \cite{Buras:2002wq}.\medskip

In the RFV framework the spurions are given by the $A$ terms. We
consider here the limiting case in which $A^u$ is flavor-diagonal so
that the CKM elements in Eq.~(\ref{eq:SoftFlavorBreaking}) are solely
generated from the $A^d_{i3}$ (``CKM generation in the down sector''). 
In this case one has\footnote{Full expressions for the effective
  Higgs couplings in RFV can be found in Ref.~\cite{RFV}.}
\begin{equation}
  \Gamma^{LR}_{sb}\,=\,\dfrac{A^d_{23}}{\mu_A}\,\propto\,V^*_{ts}\,,\hspace{2cm}
  \Gamma^{RL}_{sb}\,=\,\Gamma^{LR*}_{bs}\,\propto\,\dfrac{A^{d*}_{32}}{\mu_A}\,\propto\,V^{R*}_{32}\,. 
\end{equation}
Here we have defined
\begin{equation}
   V^R_{23}\,=\,-V^{R*}_{32}\,\equiv\,c\,\dfrac{A^{d*}_{32}}{\mu_A}\,\dfrac{v_d}{m_b}\,,\hspace{2cm}
   c\,=\,-\dfrac{2\alpha_s}{3 \pi} m_{\tilde{g}} \,\mu_A\, C_0\left(m_{\tilde{g}}^2, m_{\tilde{b}_{L}}^2,m_{\tilde{q}_{R}}^2 \right)
\end{equation}
with $m_{\tilde{b}_{L},\tilde{t}_L}$ denoting the common mass of the
left-handed sbottom and stop. The introduction of the quantity
$V^R_{32}$ simplifies the notation and allows for an easy comparison
with the size of $V_{ts}$. Since $A^d_{32}$ is a free parameter of the
theory, $\Gamma^{LR}_{bs}$ is not related to $\Gamma^{LR}_{sb}$ in RFV
and in particular not suppressed with respect to the latter. Therefore
Higgs double penguins can have sizable effects in $B_s-\bar{B}_s$
mixing, even in the light of present bounds on $\Gamma^{LR}_{sb}$,
$\Gamma^{LR}_{bs}$ from $B_s\to\mu^+\mu^-$.\medskip

\begin{figure}
\centering
\includegraphics[width=0.45\textwidth]{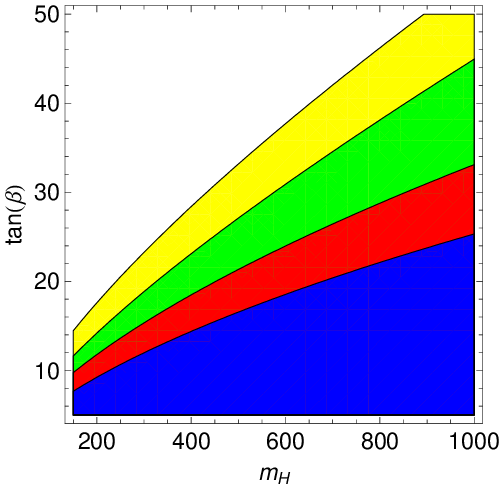}\hspace{0.08\textwidth}
\includegraphics[width=0.45\textwidth]{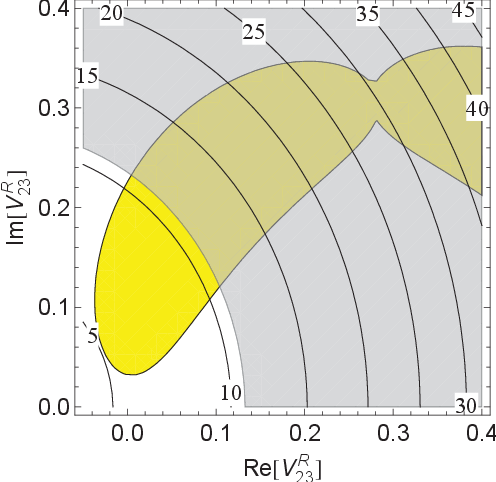}\vspace{-0.01\textheight}
\caption{Left: Allowed region in the $m_{H}$--$\tan\beta$ plane for different
  values of $\epsilon_b$ from {$\rm{Br}[B_s\to\mu^+\mu^-]\leq 1.08\cdot
  10^{-8}$}. 
  Yellow: $\epsilon_b=0.005$, green: $\epsilon_b=0.01$, 
  red: $\epsilon_b=-0.005$, 
  blue: $\epsilon_b=-0.01$ (light to dark). \newline
Right: Correlation between $B_s\to \mu^+\mu^-$ and $B_s-\bar{B}_s$ mixing for 
$\epsilon_b=0.0075$, $m_H=400\rm{GeV}$ for $\tan\beta=11$. 
Yellow: Allowed region from $B_s-\bar{B}_s$ mixing (95\% confidence
level). The contour-lines show $\rm{Br}[B_s\to
\mu^+\mu^-]\times10^9$. The grey area at the right side is excluded by the
bound on $\rm{Br}[B_s\to\mu^+\mu^-]$.\label{Bs-mixing2}}
\end{figure}

In {Fig.~\ref{Bs-mixing2}} on the left we show the allowed regions in
the $m_{H}$--$\tan\beta$ plane from $B_s\to\mu^+\mu^-$ for different
values of $\epsilon_b$. On the right the correlation between
$B_s-\bar{B}_s$ mixing and $B_s\to \mu^+\mu^-$ is shown for $m_H=400
\rm{GeV}$, $\epsilon_b=0.0075$ and $\tan\beta=11$.  Note that there is a
region in parameter space which can explain a potential new phase in
$B_s-\bar{B}_s$ mixing and which is compatible with the current limits
on $\rm{Br}[B_s\to\mu^+\mu^-]$. Moreover, if the hints for a sizable
new-physics contribution to $B_s-\bar{B}_s$ mixing persist,
$B_s\to\mu^+\mu^-$ will necessarily be enhanced. LHCb will be able to
probe this correlation in the near future.\medskip

If the CKM elements are generated from the $A_{i3}^u$ terms (``CKM
generation in the up sector''), interesting effects occur in the rare
decays $K\to \pi \nu \overline{\nu}$ (see Ref.~\cite{RFV} for details).

\subsection*{Acknowledgements}
We thank the organizers for the possibility to present our work at the EPS conference. This work is supported by BMBF grant 05H09VKF. A.C.\ acknowledges the financial support by the Swiss National Foundation.  The Albert Einstein Center for Fundamental Physics is supported by the ``Innovations- und Kooperationsprojekt C-13 of the Schweizerische Universit\"atskonferenz SUK/CRUS''. L.H. has been supported by the Helmholtz Alliance "Physics at the Terascale".


\begin{thebibliography}{99}
\bibitem{MFV}
  G.~D'Ambrosio, G.~F.~Giudice, G.~Isidori, A.~Strumia,
  Nucl.\ Phys.\  {\bf B645 } (2002)  155-187.
  [hep-ph/0207036].
  
\bibitem{Abazov:2011yk}
  V.~M.~Abazov {\it et al.} [ D0 Collaboration ],
  Phys.\ Rev.\  {\bf D84 } (2011)  052007.
  [arXiv:1106.6308 [hep-ex]]. A.~Lenz {\it et al.},
   Phys.\ Rev.\  D {\bf 83} (2011) 036004
   [arXiv:1008.1593 [hep-ph]].

\bibitem{Altm}
   W.~Altmannshofer, M.~Carena, S.~Gori, A.~de la Puente,
  [arXiv:1107.3814 [hep-ph]]. W.~Altmannshofer, M.~Carena,
  [arXiv:1110.0843 [hep-ph]].


\bibitem{U2MFV}
  R.~Barbieri, G.~Isidori, J.~Jones-Perez, P.~Lodone, D.~M.~Straub,
  Eur.\ Phys.\ J.\  {\bf C71 } (2011)  1725.
  [arXiv:1105.2296 [hep-ph]].

\bibitem{Borzumati:1999sp}
  W.~Buchmuller and D.~Wyler,
  Phys.\ Lett.\  B {\bf 121} (1983) 321.
  F.~Borzumati, G.~R.~Farrar, N.~Polonsky, S.~D.~Thomas,
  Nucl.\ Phys.\  {\bf B555 } (1999)  53-115.
  [hep-ph/9902443].
  J.~Ferrandis and N.~Haba,
  Phys.\ Rev.\  D {\bf 70} (2004) 055003
  [arXiv:hep-ph/0404077].



\bibitem{RFV}
  A.~Crivellin, L.~Hofer, U.~Nierste, D.~Scherer,
  Phys.\ Rev.\  {\bf D84 } (2011)  035030.
  [arXiv:1105.2818 [hep-ph]].
  
\bibitem{Crivellin:2010ty}
  A.~Crivellin, J.~Girrbach and U.~Nierste,
  Phys.\ Rev.\  D {\bf 83} (2011) 055009
  [arXiv:1010.4485 [hep-ph]].

\bibitem{Crivellin:2010er}
  A.~Crivellin,
  Phys.\ Rev.\  D {\bf 83} (2011) 056001
  [arXiv:1012.4840 [hep-ph]].
  A.~Crivellin, L.~Hofer and J.~Rosiek,
  JHEP {\bf 1107} (2011) 017
  [arXiv:1103.4272 [hep-ph]].

\bibitem{Gorbahn:2009pp}
  M.~Gorbahn, S.~Jager, U.~Nierste, S.~Trine,
  Phys.\ Rev.\  {\bf D84 } (2011)  034030.
  [arXiv:0901.2065 [hep-ph]].


\bibitem{Buras:2002wq}
  A.~J.~Buras, P.~H.~Chankowski, J.~Rosiek, L.~Slawianowska,
  Phys.\ Lett.\  {\bf B546 } (2002)  96-107.
  [hep-ph/0207241].

\bibitem{HNS}
  L.~Hofer, U.~Nierste, D.~Scherer,
  JHEP {\bf 0910 } (2009)  081.
  [arXiv:0907.5408 [hep-ph]].
  
\bibitem{Bsmumu}
  LHCb and CMS,
	http://inspirehep.net/record/925180/files/BPH-11-019-pas.pdf
\end{thebibliography}
\end{document}